\def\be{\begin{equation}}
\def\ee{\end{equation}}
\def\ba{\begin{eqnarray}}
\def\ea{\end{eqnarray}}
\def\l{\label}
\def\n{\nonumber \\}
\def\b{\bibitem}

\documentclass[12pt]{article}
\usepackage{epsfig}
 \begin{document} 

\title{Hard diffraction in lepton-hadron and  hadron-hadron
collisions\thanks{presented at DIS 2002, Krakow, May 2002}}

\author{A.Bialas \\ M.Smoluchowski Institute of Physics \\Jagellonian
University, Cracow\thanks{Address: Reymonta 4, 30-059 Krakow, Poland;
e-mail:bialas@th.if.uj.edu.pl;}\\and\\ H. Niewodniczanski 
Institute of Nuclear Physics, Cracow}
\maketitle

\begin{abstract} It is argued that the breakdown of factorization
observed recently in the diffractive dijet production in deep inelastic
lepton induced and hadron induced processes is naturally explained in
the Good-Walker picture of diffraction dissociation. The explicit
formula for the hadronic cross-section is given and successfully
compared with the existing data.

\end{abstract}

{\bf 1.} Diffractive production of hard jets has been recently mesured
by the CDF collaboration \cite{cdf}. When compared with the hard
diffraction observed earlier at HERA \cite{h1,zeus}, these measurements
revealed a dramatic violation of Regge factorization. The measured
diffractive structure function is about one order of magnitude smaller
than that predicted from factorization \cite{cdf, dur2,sch}. 

Two mechanisms were invoked to explain this discrepancy between the data
from (virtual)photon-induced and hadron-induced diffraction. 

The first one \cite{bhi,dur} explains the reduction of the diffractive
cross-section in hadron-induced processes by the exchange of "soft"
gluons carrying colour and thus destroying the rapidity gap (which
defines
-experimentally - the diffractive dissociation). Consequently, the
original result must be multiplied by a "gap survival probability" which
measures the probability that no soft gluon was exchanged between the
colliding particles.

In the second mechanism the "Pomeron flux" (which cannot be uniquely
defined in Regge theory) is renormalized when the incident photon is
replaced by the proton (to  prevent violation of the unitarity
condition) \cite{gul}.

In the present note I would like to suggest that

(a) The observed effect can be understood in terms of the Good-Walker
picture \cite{gw} in which the diffractive dissociation is treated as a
consequence of absorption of the particle waves\footnote{Another version
of this idea (rather different from the one presented here) was recently
discussed in \cite{dur2}.}.

(b) The magnitude of the factorization breaking can be {\it
quantitatively} estimated from the data on proton-proton elastic
scattering.

\vspace{.2cm}

{\bf 2.} In the Good-Walker formulation of diffraction dissociation the
incident particle state $|\psi>$ is expanded into a complete orthonormal
set of "diffractive eigenstates" $|\psi_n>$ which are eigenstates of the
scattering operator $T$:
\ba
T|\psi_n>= t_n|\psi_n>  \l{1}
\ea
where the eigenvalues $t_n$ are positive numbers\footnote{We use the
convention in which the high-energy elastic amplitudes (in impact
parameter representation) are real.}, not greater than 1.

To calculate the amplitude for the transition from the incident state
$|\psi>$ to a final state $|\psi'>$ (orthogonal to $|\psi>$) one expands
also $|\psi'>$ into the set $|\psi_n>$. Then the amplitude for the
transition from $|\psi>$ to $|\psi'>$ can be expressed in terms of the
expansion coefficients and the eigenvalues $t_n$.

This relation takes a particularly simple form \cite{ross,bc} if the
expansion of the observed states into the diffractive states is 
 quasi-diagonal, i.e. if we consider only  small quantum fluctuations: 
\ba |\psi>=
|\psi_1> + \epsilon |\psi_2>+...\;\;;\;\;\;\ |\psi'>= -\epsilon^* |\psi_1> +
|\psi_2> +... \l{4} 
\ea 
where $\epsilon$, the probability amplitude for the fluctuation,
 is a small number (we shall neglect
$\epsilon^2$)\footnote{$\dots$ denote other posssible small
 terms of the order $\epsilon$. They do not affect our argument.}.
 The relation between the expansion
coefficients of $|\psi>$ and $|\psi'>$ follows from the orthogonality
condition.

 Using (\ref{4}) we  obtain (keeping only  the terms linear in $\epsilon$)
\ba <\psi'|T|\psi> =\epsilon\left(t_2-t_1\right)= \epsilon
\left(<\psi_2|T|\psi_2>-<\psi_1|T|\psi_1>\right) = \n=
\epsilon
\left(<\psi'|T|\psi'>-<\psi|T|\psi>\right) . \l{5} 
\ea 
This
formula, discussed in a similar context already some time ago
\cite{ross, bc}, is the starting point of our further discussion.

To give a definite physical meaning to the Good-Walker picture we have
to define the diffractive eigenstates. Following \cite{vh} (see also
\cite{dur2, bp}) we assume that the diffractive eigenstates are states
with a fixed parton number and configuration in the transverse (impact
parameter) space. This is a natural choice since the partons, being
elementary, cannot be excited and, at high energy, their transverse
configuration is expected to remain unchanged during the collision.

\vspace{.2cm}

{\bf 3.} Consider first the photon-induced reaction:
 $|\gamma^*> \rightarrow |jets>$. We write
\ba
|\gamma^*>=|0>+\epsilon|hard>;\;\;\;\;\; |jets>=-\epsilon^*|0>+|hard>
\l{6}
\ea
where $|0>$ denotes the state with no partons and $|hard>$  a state
containing  some hard partons (decaying into  the large transverse momentum
jets in the final state).

Substituting (\ref{6}) into (\ref{5}) we obtain \ba <jets|T|\gamma^*>=
\epsilon\left(<hard|T|hard>-<0|T|0>\right) = \epsilon <hard|T|hard>
\l{7} \ea because $<0|T|0>=0$. Eq (\ref{7}) is well known since the
early discussion of vector dominance model \cite{gy}. One sees that it
can be interpreted in the Regge language: the elastic amplitude
$<hard|T|hard>$ represents the "Pomeron exchange" and $\epsilon$ is the
corresponding coupling\footnote{Note, however, that (\ref{7}) is more
general: $<hard|T|hard>$ represents the full elastic amplitude, so it
may contain the exchange of any number of Pomerons. Note also that,
unlike the standard Regge formula, (\ref{7}) is written in the impact
parameter space.}.

\vspace{.2cm}

{\bf 4.} Consider now the production of jets in diffractive proton-proton
collisions\footnote{The same argument applies for any hadron-hadron
collision.}, i.e. the transition $|P> \rightarrow |P'+jets>$, where
$|P>$ denotes the incident proton and $|P'+jets>$ contains the soft
proton remnants ($P'$) and hard jets observed in the final state.

We thus write
\ba
|P>= |soft> + \epsilon |soft'+hard>; \;\;\; \n |P'+jets>= 
-\epsilon^*|soft> +  |soft'+hard> .  \l{8}
\ea
When introduced into (\ref{5}) this gives
\ba
<P'+jets|T|P>= \epsilon\left(<soft'+hard|T|soft'+hard>-
<soft|T|soft>\right).     \l{9}
\ea
To exploit this formula we have to estimate the elastic amplitudes 
 in the R.H.S. To this end we first find that up to first
order in  $\epsilon$
\ba
<soft|T|soft> = <P|T|P>   \l{10}
\ea
To estimate $<soft'+hard|T||soft'+hard>$ we observe that it can be
treated as amplitude for scattering of a system composed of two objects:
the {\it soft} partons from the incident proton and the {\it hard} partons
which decay into the observed final jets. Thus it seems reasonable to
apply the Glauber prescription \cite{gl} and write\footnote{This idea
was already proposed in \cite{bc}.}
\ba
<soft'+hard|T|soft'+hard>= <soft'|T|soft'>+\n+<hard|T|hard>
-<hard|T|hard><soft'|T|soft'>.  \l{11}
\ea
Assuming, furthermore that 
\ba
<soft'|T|soft'> \approx <soft|T|soft>  \l{12}
\ea
we see that the soft amplitudes  in   (\ref{9}) cancel and  we
obtain
\ba
<P'+jets|T|P>= \epsilon <hard|T|hard>\left(1-<P|T|P>\right)    \l{13}
\ea
where we have used  (\ref{10}).

When compared to (\ref{7}), this formula explains the breakdown of the
factorization between the (virtual)photon-induced and hadron-induced
processes. The factor $(1-<P|T|P>)$ is usually interpreted as
"absorption" of the initial state particles. One sees, however, from its
derivation that it is actually a result of rather subtle cancellations
between the interactions in the inital and final states.

\vspace{.2cm}

{\bf 5.} Using (\ref{5}) and the formula for $(2\times 2)$ scattering
\cite{cm}, it is also not difficult to calculate the result for the
process of double diffraction dissociation. It reads
\ba
<P'_L+J_L,P'_R+J_R|T|P_L,P_R>= \epsilon_L \epsilon_R[1-<P|T|P>]
\n \left[1-(1-J_L)(1-J_R)(1-J_{LR})\right]   \l{19}
\ea
where the subscripts $(L,R)$ denote left-moving and right-moving
objects. $J_L (J_R)$ is the elastic amplitude for scattering of the
left(right)-moving {\it hard} jet system on the right(left)-moving
proton, and $J_{LR}$ is the elastic amplitude for scattering of the the
left-moving {\it hard} jet system on the right-moving one. This formula
is fairly complicated but it can be substantially simplified by
observing that the {\it hard} jet systems are represented by small size
dipoles (because of large transverse momenta of the jets) and thus the
corresponding elastic amplitudes are expected to be small. In the first
approximation (i.e. neglecting $J_{LR}$ and the higher powers of $J_L$
and $ J_R$) one obtains
\ba
<P'_L+J_L,P'_R+J_R|T|P_L,P_R> \approx \epsilon_L \epsilon_R[1-<P|T|P>]
(J_L+J_R)   \l{20}
\ea
For the {\it symmetric }
 situation (and using the notation of the previous section)
 we thus have 
\ba
<P'_L+J_L,P'_R+J_R|T|P_L,P_R> \approx 2 \epsilon^2 <hard|T|hard> [1-<P|T|P>]
   \l{21}
\ea

Comparing this with (\ref{7}) and (\ref{13}) one sees that the breaking
of factorization should be about four times less effective in the double
diffraction dissociation than the single one\footnote{The factor $2$ in
the amplitude becomes $4$ in the cross-section.}. This result seems not
too far from the recent experimental findings \cite{cdf2}.

\vspace{0.2cm}

{\bf 6.} To estimate the size of 
the discussed  effect we have taken the elastic $pp$ amplitude
in the form suggested in \cite{dl}
\ba
<P|T|P>\equiv F(t)=
\frac{\sigma_{tot}}{8\pi^2} \frac{\exp[.25t\log(s/4)]}{(1-t/.71)^4}
\l{15}
\ea
from which one can calculate the impact parameter representation needed
in (\ref{13}). 
The product  $\epsilon \times <hard|T|hard>$ was taken as a Gaussian
\ba
<hard|T|hard> \sim exp(-b^2/2B)  \l{16}
\ea
where $B$ is the slope of the   {\it cross-section} in the
 (virtual)photon-induced process  (\ref{5}).

The hadron-induced diffraction dissociation {\it cross-section} 
can then be expressed as 
\ba \sigma(P\rightarrow P'+jets) = R
\;\;\sigma_{factorized}(P\rightarrow P'+jets) \l{17} 
\ea 
where
$\sigma_{factorized}$ denotes the cross-section extrapolated from the
deep inelastic scattering data, and 
\ba R= 1- 2 \pi \int dt \exp(tB/4)
F(t) +\n+ \pi^2 \int dt dt' \exp(tB/4) F(t)\exp(t'B/4) F(t')
I_0(\sqrt{tt'}B/2) \l{18} 
\ea

This expression depends on one unknown parameter, B - the slope in the
momentum transfer dependence of the diffractive jet production in deep
inelastic scattering. For production of heavy vector mesons $B \approx 4
$ GeV$^{-2}$ \cite{cr}. One can speculate that this is a lower limit for
$B$ which may be approximately valid for production of jets with a small
mass (large $\beta$). As the mass increases ($\beta$ decreases), one may
expect that $B$ should increase (the system becomes more complicated and
its transverse size is expected to grow)\footnote{This is confirmed by
the measurements of the {\it inclusive} diffraction at HERA where one
finds $B\approx$ 7 GeV $^{-2}$ \cite{B}.}. 
\begin{figure}[hbt]
\centerline{%
\epsfig{file=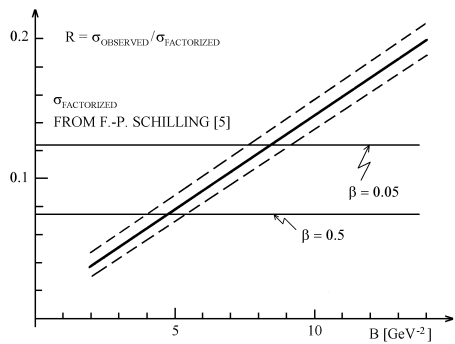,width=6cm}}
\caption{\small Ratio $R$ plotted versus $B$. The horizontal lines 
represent the phenomenological estimates of $R$ given in \cite{sch}.}
\end{figure}

In Figure 1 we show, plotted versus $B$, the ratio $R$ calculated
from (\ref{18}), using $\sigma_{tot}=71.7 \pm 2 mb$ \cite{st}. The
recent phenomenological estimates of $R$, given in \cite{sch} are also
shown. One sees that the result is certainly not far from the data.

\vspace{0.2cm}

{\bf 6.} Some comments are in order. 

(i) One sees from the discussion in Section 3 that the {\it uncorrected}
formula (\ref{7}) is valid independently of the virtuality of the
incident photon: The same formula applies to photoproduction and to deep
inelastic scattering. This emphasizes the (already mentioned) point: the
effect we consider cannot be simply identified with absorption in the
initial state of the process.

(ii) Using the cross-sections at other energies, one can investigate the
energy dependence of the correction factor $R$. Taking
$\sigma_{tot}$(630)= 63 mb, one finds that $R(630)/R(1800)$ varies from
$\sim 1.5$ ($B=4$ GeV$^{-2}$) to $\sim 1.2$ ($B=10$ GeV$^{-2}$), in a
reasonable agreement with recent data from the CDF Collaboration
\cite{cdf3}.

(iii) In the numerical estimate of Section 5 we have assumed that the
dipole corresponding to the two jets is created at the same impact
parameter as the incident proton. This assumption seems rather 
natural\footnote{As long as the diffractive system is produced in the proton
vertex. If, however, a large rapidity gap develops (i.e. for jet
production in the central rapidity region, corresponding to "double
Pomeron exchange" processes) one may expect that the produced system is
far from the proton remnants in the impact parameter space. In this case
the ratio $R$ would be close to 1.} but some deviations  cannot be
excluded. They would  increase somewhat the correction factor $R$.

(iv) Our result given in Eq.(\ref{13}) resembles, to some extent, the
"renormalization" of the Pomeron flux, proposed in \cite{gul}. One
should keep in mind, however, that the Eq. (\ref{13}) refers to impact
parameter space and thus it can be at best only approximately
interpreted as the (corrected) Regge formula. 

(v) It is not unlikely that an argument similar to the one presented
here can be also applied to the {\it soft} diffraction dissociation. It
would be certainly very interesting to analyze the data from this point
of view.

\vspace{0.2cm}

{\bf 7. } In conlusion, we have shown that the breakdown of Regge
factorization between the diffractive production of hard jets observed
at HERA and at FERMILAB is naturally explained in the Good-Walker
picture of diffraction dissociation. The correction to the factorization
formula is explicitely given in terms of the elastic $p\bar{p}$
amplitude at small momentum transfers. The numerical estimates seem to
be consistent with the experimental findings.

\vspace{0.3cm}
{\bf Acknowledgements}
\vspace{0.3cm}

I greatly profited from discussions with W.Czyz, K.Goulianos, K.Fialkowski, 
 A.Kotanski and R.Peschanski.
This investigation was supported in part by the 
 Subsydium of Foundation for Polish Science NP 1/99 and by the KBN
grant No 2 P03 B 09322.

\end{document}